\documentclass[intlimits,twoside,a4paper]{article}

\usepackage[utf8]{inputenc}

\usepackage[eqsecnum]{cmpj3}
\issue{2024}{27}{4}{43703}
\doinumber{10.5488/CMP.27.43703}
\title[Electron and hole energy spectrum of non-concentric spherical core--shell quantum dot]%
{Electron and hole energy spectrum of non-concentric spherical core--shell quantum dot under an externally applied electric field}
\author[R. Ya. Leshko, I. V. Bilynskyi, O. V. Leshko, M. Yu. Popov, A. O. Ocheretyanyi]
{R.~Ya.~Leshko\orcid{0000-0002-9072-164X}\refaddr{label1}\thanks{Corresponding author: \email{leshkoroman@gmail.com}.}, 
        I.~V.~Bilynskyi\orcid{0000-0002-4221-9225}\refaddr{label1,label2}, 
		O.~V.~Leshko\orcid{0000-0001-9646-3189}\refaddr{label1}, 
		M.~Yu.~Popov\refaddr{label2}, 
		A.~O.~Ocheretyanyi\refaddr{label2}
}
\addresses{
\addr{label1} Department of Physics and Information Systems, Drohobych Ivan Franko State Pedagogical University, 3 Stryiska Street, 82100 Drohobych, Ukraine
\addr{label2} Department of Physics, Kryvyi Rih State Pedagogical University, 54 Universytetska Street, 50086 Kryvyi Rih, Ukraine
}

\Keywords{non-concentric spherical core--shell quantum dot, electric field}

\date{Received November 11, 2024, in final form December 06, 2024}

\begin{document}

\maketitle

\begin{abstract}
	A model of the non-concentric spherical core--shell quantum dot under the influence 	of an externally applied electric field was proposed. 
	It was established that the energy spectrum of both the electron and the hole depends 	on the intensity of the electric field as well as on the specific location of the core within the quantum dot. The phenomenon of energy level splitting and degeneration was analyzed in detail. Additionally, the variations in the optical gap were determined and expressed as a function of the applied electric field strength and 
	the position of the core in the quantum dot.
%
%
%\keywords Up to six keywords (\href{https://physh.aps.org/browse}{Physics Subject Headings})
\printkeywords
%
%\pacs Up to six PACS numbers (optional)
\end{abstract}

\section{Introduction}

%\doclicenseThis

In recent years, significant advancements have been achieved in the field of nanotechnology, 
particularly in the production of hybrid quantum dots (QDs) composed of multiple components. 
By combining different materials, it has become possible to fabricate multilayer QDs with 
diverse and tunable properties. This integration of various components has opened up new 
opportunities for enhancing the performance and efficiency of optoelectronic devices because the use of multilayer QDs can improve the operational parameters of these devices. 
This approach allows for expanded functionality and customization of optoelectronic 
systems in order to meet specific requirements.

Spherical core--shell quantum dots (CSQD) are the most basic type of multilayer structures. 
The interest in CSQDs stems from their capability to modify the essential optical 
properties of the core nanocrystals, such as the fluorescence emission wavelength and quantum yield by varying the thickness of the shell surrounding the core \cite{1,2,3,4}. 
In these structures, due to the presence of the shell, a reduced blinking is observed \cite{5,6}. 
Those and other advantages caused the rise of theoretical interest in studying CSQDs 
(for example \cite{7,8,9,10,11,12,13,14,15,16}).
Electron and hole spectra were calculated in \cite{7,8,9,10,11}, 
taking into account external electric fields \cite{8,9,10} and a magnetic field \cite{8}.
Based on the electron and hole spectra, the photoionization cross-section \cite{9,12}, interlevel, 
and intersubband transition energies \cite{13,14} were obtained, 
and the linear and nonlinear optical properties 
of CSQDs were calculated. Furthermore, the influence of QD-matrix 
deformation on the baric properties of CSQDs 
was determined \cite{16}. In works~\cite{17,18,19} 
it was shown that photoluminescence flickering can be controlled 
by an external action, in particular, an electric field.

The above mentioned works further confirm that the study of the effect of the electric field on the CSQD 
is an actual task that is solved in many works. However, in most of them, 
the concentric spherical CSQDs are considered. 
In real situations, there is no certainty that the spherical CSQD will be concentric. 
Furthermore, the violation of concentricity can be reached by cation exchange in layers \cite{20}. 
In work~\cite{20}, high-resolution transmission electron microscopy and high-angle annular dark-field 
scanning transmission electron microscopy images are presented, which prove that 
non-concentric spherical CSQDs were obtained. 
There are presented numerical calculations of electron and hole spectrum 
in Comsol (software) using the finite element method. 
However, there is no analytical theory of non-concentric spherical CSQDs. 
The first analytic model of non-concentric CSQD was presented in our work \cite{21}.

Calculations \cite{20,21} show that the core displacement caused the changes in the hole and electron levels. 
The energy levels of the excited states split \cite{21}. 
This splitting can be changed depending on the direction and value of the  electric field. 
It can increase the changes in electron spectra caused by the core displacement and decrease them, 
like we had in \cite{22} with electric field and impurity. 
The external electric field will change the hole levels and the effective optical gap.

For this reason, the objective of this paper is to 
investigate the energy states of both electrons 
and holes in a non-concentric CSQD exposed to an external electric field. 
It also aims at assessing the impact of the position of the core, 
as well as of the magnitude and direction of the electric field on the effective optical gap.

\section{Theory}

\begin{figure}[htb]
	\centerline{\includegraphics[width=0.55\textwidth]{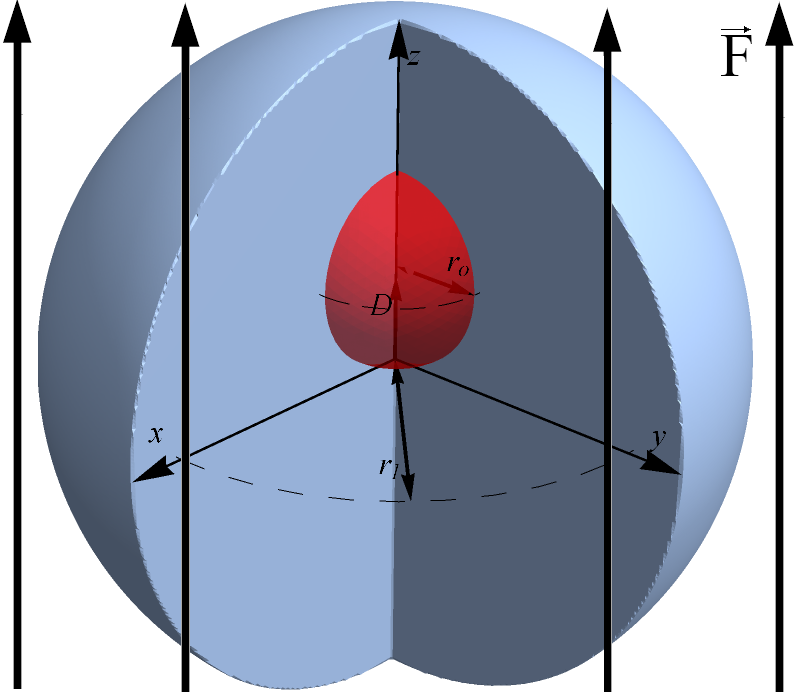}}
	\caption{(Colour online) Geometric model of core--shell QD.} \label{fig1}
\end{figure}

Let us consider the semiconductor spherical non-concentric core--shell QD.
The core radius is $r_0$, the shell radius is $r_1$.
The QD is characterized by the effective electron mass: 
$m_0$ in the core, $m_1$ in the shell. 
The CSQD is in the bulk matrix with the electron effective mass $m_2$.
Additionally, we assume that the lattice constants and 
dielectric constants of materials have close values,
which is why we neglect the impact of polarization and deformation effects.
We use an average value of the dielectric permitivity $\varepsilon$. 
In our analysis, we focus on the case where the QD core is displaced 
from the QD center by a distance $D$ in the $z$-direction 
(as illustrated in figure~\ref{fig1}), where $r_1-r_0 \leqslant D$.
The external electric field $\vec{F} = (F_x, F_y, F_z)$ is applied to the heterosystem. 
Electrostatic potential energy of the electron in external field has the following form
\begin{equation}
	\label{eq1}
	V(\vec{r}) = - \vec{d} \cdot \vec{F} = e \vec{r} \cdot \vec{F} = e(xF_x +yF_y + zF_z),
\end{equation}

\noindent where $\vec{d}$ is dipole momentum, $e$ is elementary charge 
($-e$ is the electron charge). We consider the case of $z$-direction of the electric field
\begin{equation}
	\vec{F} = (0,0,F).
	\label{eq2}
\end{equation}
The electron Hamiltonian of the described system in units of 
effective Rydberg energy $\textrm{Ry}^{*} = \frac{\hbar^2}{2m_0a_b^{*2}}$
and effective Bohr radius $a_b^{*} = \frac{\hbar^2 \varepsilon}{m_0 e^2}$ is as follows:
\begin{equation}
	\hat{H} = - \vec{\nabla} \cdot \left[\frac{m_0}{m(\vec{r})} \vec{\nabla} \right] + U(\vec{r}) + \tilde{V}(\vec{r}),
	\label{eq3}
\end{equation}
\noindent where
\begin{align}
	\label{eq4}
		m\left( \vec{r} \right) = \left\{ \begin{array}{l}
		{m_0},\quad\vec{r} \quad \text{in core},\\
		{m_1},\quad \vec{r} \quad \text{in shell}, \\
		{m_2},\quad \vec{r} \quad \text{in matrix}\\
		\end{array} \right.
\end{align}
\noindent is the electron effective mass,
\begin{align}
	\label{eq5}
		U\left( \vec{r} \right) = \left\{ \begin{array}{l}
		{0},\qquad \vec{r} \quad \text{in core},\\
		{U_{01}},\quad \vec{r} \quad \text{in shell}, \\
		{U_{02}},\quad \vec{r} \quad \text{in matrix} \\
		\end{array} \right.
\end{align}
\noindent is confinement potential, and $\tilde{V}(\vec{r}) = V(\vec{r})/\textrm{Ry}^{*}$. 
Hereafter, we omit the tilde symbol and assume that the distances 
are measured in units of the effective Bohr radius, 
while the energies are measured in units of the effective Rydberg energy.

In order to determine the energy spectrum and wave functions of the electron, 
one must solve the Schr\"{o}dinger equation. 
However, when dealing with the systems that lack a spherical symmetry, 
exact solutions to this equation are not feasible. 
Hence, the plane wave method is employed, 
which is extensively explained in references \cite{21, 23, 24, 25}.
The wave function, serving as a solution to the Schr\"{o}dinger 
equation, can be represented in the following manner:
\begin{equation}
	\label{eq6}
	\psi(\vec{r}) = \sum_{n_x, n_y, n_z}^{} {C_{n_x, n_y, n_z} \psi^{(0)}_{n_x, n_y, n_z}(x, y, z)},
\end{equation}
\noindent where 
\begin{equation}
	\label{eq7}
	\psi^{(0)}_{n_x, n_y, n_z} = \frac{1}{\sqrt{L_x L_y L_z}}
		\exp \left\{ {
			\ri \left[ 
				(k_x + n_x K_x)x + (k_y + n_y K_y)y + (k_z + n_z K_z)z
			\right]
		} \right\}.
\end{equation}
\noindent Here, $L_x = L_y = L_z \equiv L$ are the edge lengths 
of the unit cell along the $x$, $y$, and $z$ directions 
of the coordinate system, $K_x = K_y = K_z \equiv 2\piup / L$,
\begin{equation}
	\label{eq8}
	{n}_{x} \in [-{n}_{\rm max}, ..., {n}_{\rm max} ], {n}_{y} \in [-{n}_{\rm max}, ..., {n}_{\rm max} ], {n}_{z} \in[-{n}_{\rm max}, ..., {n}_{\rm max} ].
\end{equation}
References \cite{21, 23, 24, 25} demonstrated that the results achieved convergence by considering 
$n_{\rm max} = 7$ and $L = 2.5 + 2 r_1 $. Furthermore, it was
substantiated that the obtained results are independent 
of the wave vector ($k_x$, $k_y$, $k_z$) when using those parameters. 
That is why, we get $k_x = k_y = k_z = 0$ in the following calculations.

Upon substituting equation (\ref{eq6}) into the Schr\"{o}dinger equation containing Hamiltonian (\ref{eq3}), 
a system of linear homogeneous equations is obtained:
\begin{equation}
	\label{eq9}
	\sum_{n_x, n_y, n_z} 
		{\left( {T_{{n'_x, n'_y, n'_z}\atop
				{n_x, n_y, n_z}}} 
			+ 
			{U_{{n'_x, n'_y, n'_z}\atop
					{n_x, n_y, n_z}}}
			+
			{V_{{n'_x, n'_y, n'_z}\atop
					{n_x, n_y, n_z}}} 
			- E {\delta_{{n'_x, n'_y, n'_z}\atop
					{n_x, n_y, n_z}}} 
		 \right) C_{n_x, n_y, n_z}} = 0.
\end{equation}
\noindent Matrix elements $T$ and $U$ were presented and derived in \cite{21}.
The matrix element of (\ref{eq1}) has the following form
\begin{equation}
	\label{eq10}
	V_{{n'_x, n'_y, n'_z}\atop
		{n_x, n_y, n_z}} = e F L \delta_{n'_x,n_x} \delta_{n'_y, n_y}
		\left\{ \begin{array}{l}
			{0},\,\,\,\,\,\,\,\,\,\,\,\,\,\,\,\,\,\,\,\,\,\,\,\,\,\,\,\,n'_z = n_z,\\
			{\frac{-\ri(-1)^{n_z-n'_z}}{2\piup(n_z - n'_z)}},\,\,\,\ n'_z \neq  n_z.
			\end{array} \right.	
\end{equation}
\noindent From the system of linear homogeneous equations (\ref{eq9}) and normalized condition, 
the energy $E$ and all coefficients can be found. In other words, we diagonalize the matrix $T+U+V$.
Therefore, the wave function (\ref{eq6})
can be defined.

For the hole states, we assume that the valence band can be described by a simple isotropic 
parabolic dispersion relation. That is why we used all the presented formulae 
(\ref{eq1})--(\ref{eq10}) for the holes with a suitable replacement of effective masses (\ref{eq4})
and confinement potential (\ref{eq5}).

\section{Calculation results}

We perform calculations for the CSQD of the heterostructure GaAs/Al$_x$Ga$_{1-x}$As/matrix using the physical parameters like in our previous works \cite{21, 22}: 
a) for the electron ($x = 0.4$, $m_0 = 0.067m_e$, $m_1 = 0.1m_e$, $m_2 = m_e$, $U_{01} = 297$~meV); 
b) for the hole ($x = 0.4$, $m_0 = 0.51m_e$, $m_1 = 0.61m_e$, $m_2 = m_e$, $U_{01} = 562$ meV).
$m_e$ is the free electron mass, $E_g = 451$~meV. 
We also consider the matrix, where the band offset between QD shell and matrix is very large.
That is why we assume that $U_{02} = 6000$~meV.

\begin{figure}[h!]
	\centerline{\includegraphics[width=0.94\textwidth]{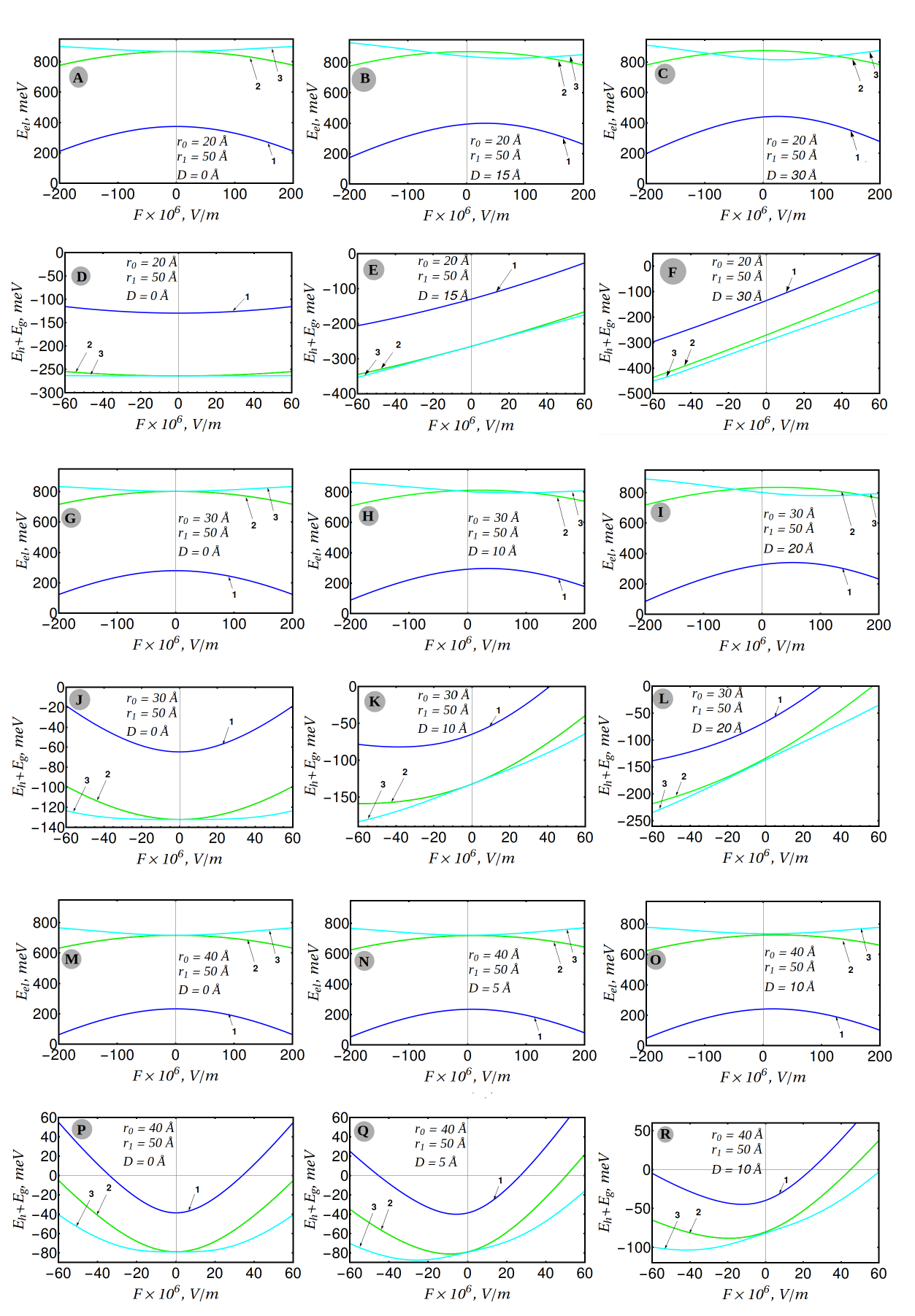}}
	\caption{(Colour online) 
		Electron (A, B, C, G, H, I, M, N, O) 
		and hole (D, E, F, J, K, L, P, Q, R) energy levels as a 
		function of the electric field. 
		1 --- the energy of ground state ($s$-levels); 
		2, 3 --- energies of exited states ($p$-levels); 
		2 --- magnetic quantum number $|m_l| = 0$; 
		3 --- $|m_l| = 1$.} 
	\label{fig2}
\end{figure}

The dependence of the energy levels of the electron and hole on the magnitude 
of the applied electric field along the $z$-axis is illustrated in figure~\ref{fig2}.
There we consider a concentric case [the first column (A, D, G, J, M, P) of graphs in figure~\ref{fig2}]; 
the case $D = {( {r}_{1} - {r}_{0} )} /{2}$ [the second column (B, E, H, K, N, Q) of graphs in figure~\ref{fig2}];
and $D = ( {r}_{1} - {r}_{0} )$ [the third column of graphs (C, F, I, L, O, R) in figure~\ref{fig2}]. 
In all these cases, $r_1 = 50$~\AA.

\begin{figure}[!h]
	\centerline{\includegraphics[width=0.8\textwidth]{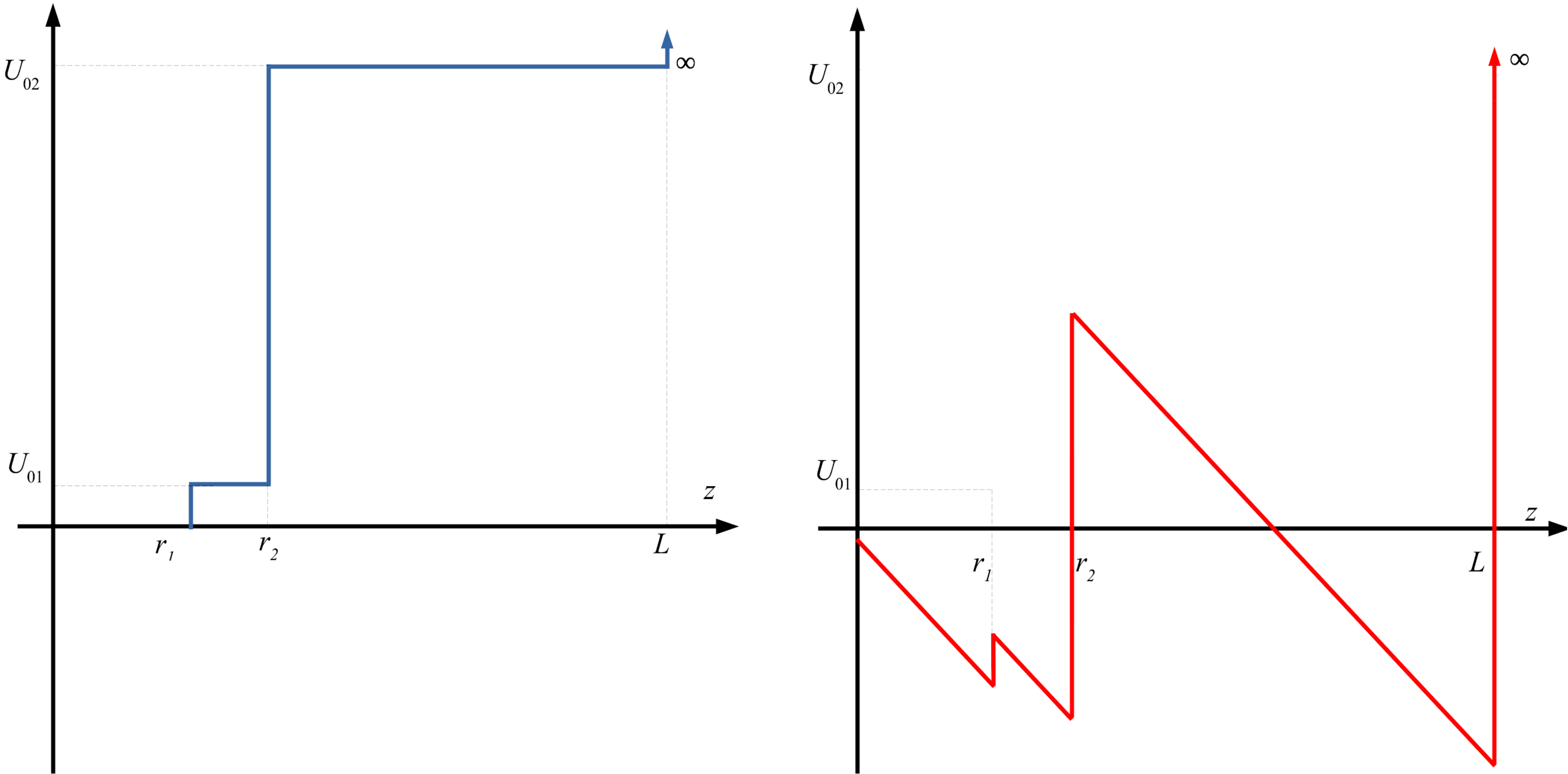}}
	\caption{(Colour online) 
		The energy model of potential wells and barriers in 
		$z$-direction of the concentric core--shell QD without 
		electric field (left-hand side; blue color) and with electric 
		field (right-hand side; red color).} 
	\label{fig3}
\end{figure}

From the graphs in figure~\ref{fig2}, it is evident that we calculated 
the dependence of the energy of the electron within the range of electric 
field values from $-60 \cdot 10^6$ to $60 \cdot 10^6$~V/m, 
while the holes were within the range of $-200 \cdot 10^6$ to $200 \cdot 10^6$~V/m.
Such a selection was dictated by the plane wave method that requires 
the existence of infinitely high potential walls at the boundaries 
of a cube with sides of length $L$. The energy model of potential 
wells and barriers is schematically depicted in figure~\ref{fig3}.

From the energy diagram (figure~\ref{fig3}), it can be observed that as the magnitude 
of the electric field increases, the zones tilted more. Therefore,
for high electric fields, conditions may arise where a particle 
with a predominant probability will reside in the region from $r_2$
to $L$, close to the surface of the cube $L$. 
Consequently, the energy levels will also be located in the same 
region (see figure~\ref{fig3}). 
Since $L$ is large, this situation corresponds to a case where 
the particle has left the QD. We will not consider these cases. 
Consequently, for the electron and the hole, there are specific 
electric field values at which such a process occurs. Therefore, 
for the radii of the core, shell, and the displacement of the 
core from the center (which are used in figure~\ref{fig2}), 
the electric field values (when the departure from the electron 
or hole from the quantum dot is not observed) are the same as mentioned above. 
As the effective masses of the holes are larger, their energy 
is consequently lower by absolute value. 
Hence, the tunneling effect into the matrix for holes can occur 
at lower electric fields.

In the case of a concentric CSQD, the applied electric 
field caused the splitting of electron and hole $1p$ levels 
(figure~\ref{fig2} A, D, G, J, M, P). 
However, in the case of non-concentric CSQD, the electron and hole levels 
are split in zero electric fields by the magnetic quantum number $|m_l|$.
The magnitude of splitting depends on the QD layers size~\cite{21}. 
Furthermore, in~\cite{21} we showed that the splitt $p$-levels can change the order. 
In the electric field, these levels can degenerate again or 
the splitting can increase.

Let us consider the electron levels in a non-concentric CSQD. 
If the electric field is directed along the $z$ axis ($\vec{F} \uparrow \uparrow  z$,  $F>0$),
the electric field ``pushes'' the electron from the shifted core 
to the shell in $-z$ direction (opposite to the direction of the 
electric field). In this case, compensatory effects occur. 
That is, at a certain value of the electric field magnitude, 
the initial spherical symmetry is restored 
(similar to the case of the concentric spherical CSQD). 
A further increase in the electric field magnitude again breaks 
this spherical symmetry. Therefore, for the ground state energy, 
as the electric field increases, we observe a maximum in the $E(F)$ 
dependence. For example, this is observed in  (figure~\ref{fig2}~I). 
The $p$ levels may have 1 or 2 points of degeneration recovery (for $D>0$). 
One point of degeneration recovery is always observed when $F>0$. 
The second point can arise at $D>0$ and $F<0$, when the $p$-level order 
is changed.

\begin{figure}[!h]
	\centerline{\includegraphics[width=0.9\textwidth]{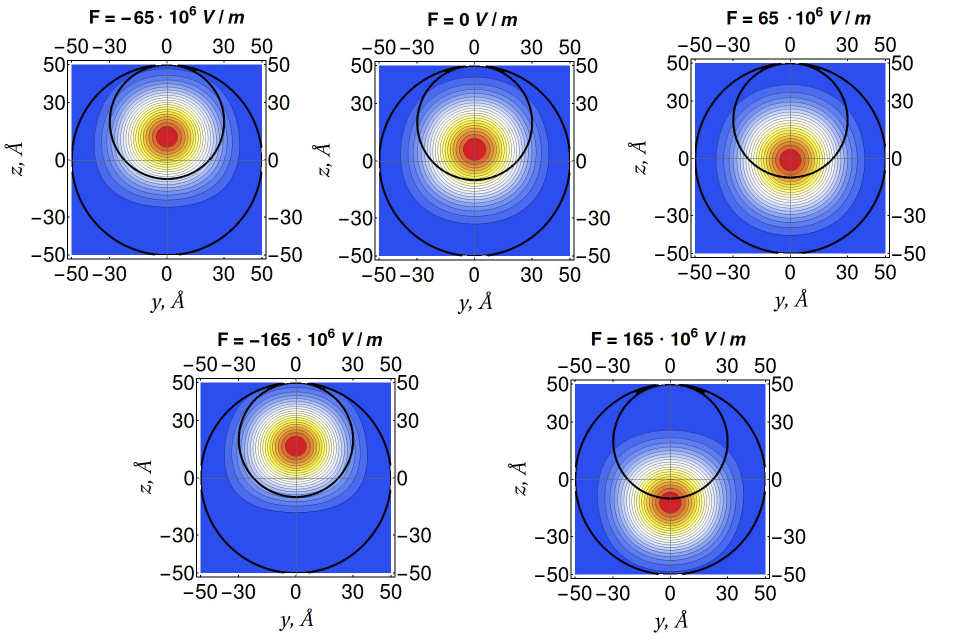}}
	\caption{(Colour online) 
		The probability density of electron location in the QD at 
		different electric fields. \\
		$r_0=30$~\AA, 
		$r_1=50$~\AA, 
		$D=20$~\AA.} 
		\label{fig4}
\end{figure}

To confirm the conclusions regarding the  restored initial spherical symmetry,
we built the graphs illustrating the distribution of electron 
probability density in the ground state for a non-concentric 
spherical CSQD at various values of the electric field (figure~\ref{fig4}). 
It can be observed that negative values 
(when the field is directed opposite to the $z$-axis) 
of the electric field cause a further displacement of the probability 
density maximum along the $z$-axis (figure~\ref{fig4}, $F > 0$). 
Whereas positive values ($F > 0$) lead to the restoration of 
spherical symmetry ($F = 65 \cdot {10} ^ {6}$~V/m) and to further 
displacement of the probability density maximum in the opposite 
direction to the $z$-axis.
Similar dependencies have be observed for the hole states too.

\begin{figure}[!h]
	\centerline{\includegraphics[width=0.9\textwidth]{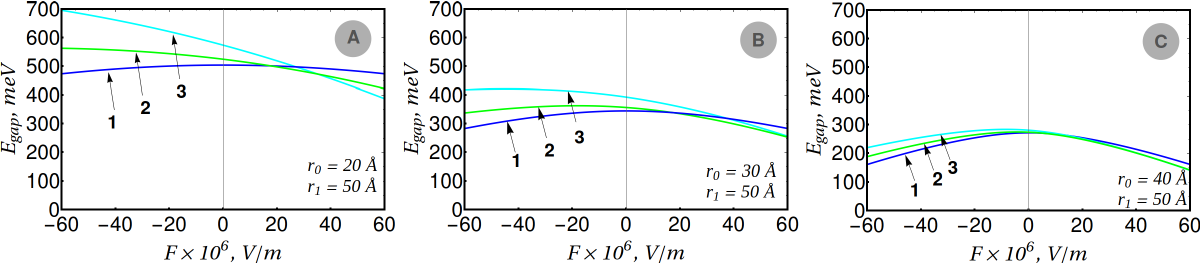}}
	\caption{(Colour online) 
		The effective optical gap (${E}_{\rm gap} = {E}_{e} - {E}_{h}-{E}_{g}$) as a function of the  
		applied electric field.  Shell radius is $r_1 = 50$~\AA.   
		Core radii are $r_0=20$~\AA \ (A), $r_0=30$~\AA \ (B), 
		$r_1=40$~\AA~(C). 
		Curve~1 corresponds to $D = 0$; curve 2 corresponds to $D = ( {r}_{1} - {r}_{0} ) / {2}$; 
		curve 3 corresponds to $D = ( {r}_{1} - {r}_{0} )$.} 
		\label{fig5}
\end{figure}

Taking into account the change in the energy of the hole and electron 
levels in a non-concentric spherical core--shell QD under an electric 
field, it is possible to determine the dependence of the effective 
optical gap on the electric field and on the magnitude of the core 
displacement. This result is presented in figure~\ref{fig5}.

From figure~\ref{fig5}, it can be seen that for a concentric 
spherical CSQD (curves 1, figure~\ref{fig5}), the application of 
an electric field reduces the effective optical gap. 
It is also seen that in the absence of an electric field ($F = 0$), 
the effective optical gap is larger in the case of a non-concentric 
CSQD. Moreover, the larger is parameter $D$ (non-concentricity parameter) 
 the larger is the optical gap (curves 1, 2, 3 in figure~\ref{fig5} 
at $F = 0$). 
The applied non-zero electric field induces a change in the magnitude 
of the optical gap. Specifically, its value increases when electronic 
levels rise and decreases when the hole levels decrease. 
Conversely, the reduction in the optical gap occurs in the opposite 
scenario.

\section{Conclusion}
The conducted computations of the electronic and hole levels in 
a spherical CSQD enabled the establishment of the dependence 
of the energy levels on the magnitude and direction of the electric field. 
It was determined that an increase in non-concentricity leads to an 
increase in the ground state energy for electrons and to a decrease 
for the holes. The energy levels of excited $p$-states split according to 
the magnetic quantum number due to the disruption of spherical 
symmetry and conservation of cylindrical symmetry.

The obtained dependencies of energies in non-concentric CSQDs 
concerning the electric field demonstrate that there are electric 
fields causing the restoration of spherical symmetry by compensating 
for the influence of non-concentricity. 
Degeneracy occurs again for the energy levels of $p$-states at 
specific electric fields too.

The obtained results can be utilized to determine the absorption 
and luminescence of non-concentric spherical CSQDs both 
in the presence and in the absence of an electric field. 
Additionally, the theory can be extended to the case of 
arbitrary direction of the electric field.

% \bibliographystyle{cmpj}
% \bibliography{cmpjxampl}

%
%% If you have problems with typesetting in ukrainian uncomment lines below.
%
%  \lastpage
%  \end{document}

\ukrainianpart

\title{Електронний і дірковий спектри неконцентричної сферичної квантової точки типу ядро-оболонка у зовнішньому електричному полі}
\author{Р.~Я.~Лешко\refaddr{label1},
 		І.~В.~Білинський\refaddr{label1, label2},
		О.~В.~Лешко\refaddr{label1},
		М.~Ю.~Попов\refaddr{label2},
		A.~О.~Очеретяний\refaddr{label2}
		}
\addresses{
\addr{label1} Кафедра фізики та інформаційних систем, Дрогобицький державний перадгогічний університет імені Івана Франка,
вул. Стрийська 3, 82100, Дрогобич, Україна
\addr{label2} Кафедра фізики, Криворізький державний педагогічний університет,
вул. Університетська 54, 50086, Кривий Ріг, Україна
}

\makeukrtitle

\begin{abstract}
\tolerance=3000%
Було запропоновано модель неконцентричної сферичної 
квантової точки типу ядро-оболонка під впливом зовнішнього 
електричного поля. Встановлено, що енергетичний спектр як 
електрона, так і дірки залежить від величини 
електричного поля, а також від конкретного розташування 
ядра всередині квантової точки. Особливості розщеплення та 
виродження енергетичних рівнів було детально проаналізовано. 
Додатково було визначено зміну ефективної оптичної щілини, 
функцію від величини прикладеного електричного поля та 
положення ядра.
\keywords неконцентрична сферична квантова точка типу ядро-оболонка, 
електричне поле

\end{abstract}

\end{document}